# Interaction energy between two separated charged spheres surrounded inside and outside by electrolyte

István P. Sugár

Department of Neurology, Icahn School of Medicine at Mount Sinai, New York, NY 10029

## Abstract

By using the recently generalized version of Newton's Shell Theorem [1] analytical equations are derived to calculate the electric interaction energy between two separated charged spheres surrounded outside and inside by electrolyte. This electric interaction energy is calculated as a function of the electrolyte's ion concentration, temperature, distance between the spheres and size of the spheres. At the same distance between the spheres the absolute value of the interaction energy decreases with increasing electrolyte ion concentration and increases with increasing temperature. At zero electrolyte ion concentration the derived analytical equation transforms into the Coulomb equation. Finally, the analytical equation is generalized to calculate the electric interaction energy of N separated charged spheres surrounded by electrolyte.

## Introduction

The head groups of membrane lipids have either single charge (e.g. tetraether lipids [2,3]) or electric dipole (e.g. phospholipids [4,5]). Theoretical models of lipid membranes usually focus on short range (Van der Waals) lateral interactions between nearest neighbor lipids and ignore the long range charge-charge interactions [5]. This is because in the case of long range interactions one has to consider the entire system rather than the lateral interactions between the nearest neighbor lipids only. In order to get closer to the solution of this problem recently we developed a generalized version of Newton's Shell Theorem [1]. According to the generalized Shell Theorem the potential around a charged sphere of radius $R$ is (see Eq.9 in ref.1)

$$V(Z, Q) = \frac{k_e Q \lambda_D}{\varepsilon_r Z R} e^{-\frac{Z}{\lambda_D}} \sinh\left(\frac{R}{\lambda_D}\right) \qquad \text{at } Z > R \qquad (1)$$

where $Q = 4R^2 \pi \rho_s$ is the total charge of the sphere and $\rho_s$ is the surface charge density, $\lambda_D$ is the Debye length in the electrolyte that is inside and around the charged sphere, $k_e$ is the Coulomb's constant and $\varepsilon_r$ is the relative static permittivity of the electrolyte. In ref.1 we also calculated the electric potential of two concentric charged spheres surrounded by electrolyte, and the membrane potential of a charged lipid vesicle surrounded by electrolyte with high ion concentration. At any electrolyte concentration one can calculate the electric potential of the charged lipid vesicle by numerical integration (see ref.1).

In this paper we consider two separated charged spheres surrounded outside and inside by overall neutral electrolyte containing only monovalent positive and negative ions and calculate its Debye length by Eq.A2. By using Eq.1 an analytical equation is derived to calculate the electric interaction energy between the charged spheres. By means of this analytical equation one can calculate the dependence of the electric interaction energy from the distance, charge and size of the spheres and from the electrolyte's ion concentration and the temperature. In the case of our calculations the surface charge density of the charged spheres at every radius is $\rho_s = -0.266\ C/m^2$. This is the charge density of PLFE (bipolar tetraether lipid with the polar lipid fraction E) vesicles if the cross sectional area of a PLFE is $0.6 nm^2$ and the charge of a PLFE molecule is $-1.6 \cdot 10^{-19} C$ [2,3].

## Model

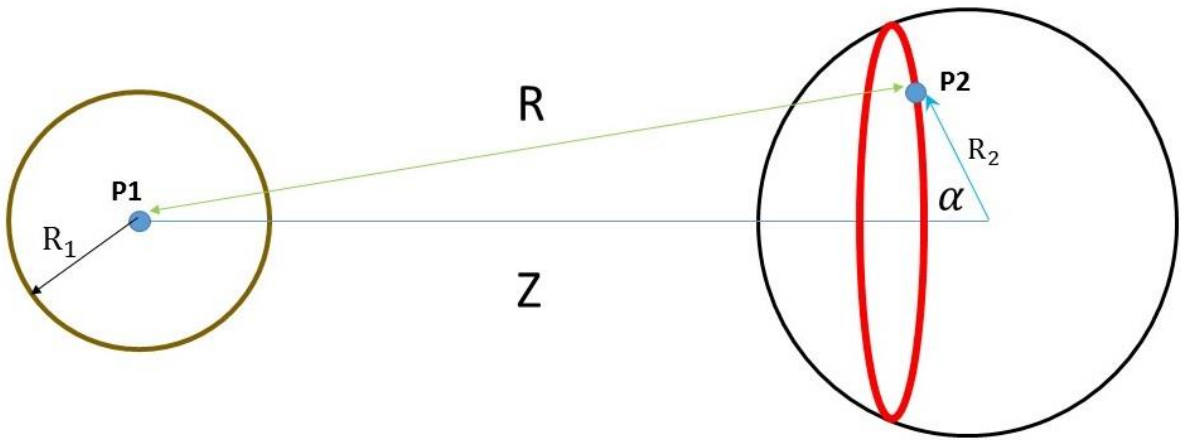

**Figure 1.** *Two charged spheres*
Left circle represents a charged sphere of radius $R_1$ and its total surface charge is $Q_1$. Right circle represents a charged sphere of radius $R_2$ and its total surface charge is $Q_2$. The distance between the centers of the two spheres is Z. The potential created by the left charged sphere is calculated at point P2.
Red ring represents charges on the right charged sphere. Their distance from point P1 is R. $\alpha$ is the angle between vector $Z$ and a vector pointing from the center of the right sphere to any of the point (P2) of the red ring.

Based on the generalized Shell Theorem [1] the electric potential created by the left charged sphere at point P2 is:

$$V(R) = \frac{k_e Q_1 \lambda_D}{\varepsilon_r R R_1} e^{-\frac{R}{\lambda_D}} sinh\left(\frac{R_1}{\lambda_D}\right) \tag{2}$$

The distance between point P1 and any of the point charges located on the red ring is:

$$R(\alpha, Z, R_2) = \sqrt{(R_2 \sin(\alpha))^2 + (Z - R_2 \cos(\alpha))^2} = \sqrt{R_2^2 + Z^2 - 2ZR_2 \cos(\alpha)} \tag{3}$$

The interaction energy between the left charged sphere and the charges of the red ring is:

$$E(\alpha)d\alpha = V(R)\rho_2 2R_2 \sin(\alpha) \pi R_2 \cdot d\alpha = \frac{k_e Q_1 \lambda_D}{\varepsilon_r R(\alpha,Z,R_2) R_1} e^{-\frac{R(\alpha,Z,R_2)}{\lambda_D}} sinh\left(\frac{R_1}{\lambda_D}\right) \frac{Q_2}{2} \sin(\alpha) \cdot d\alpha \tag{4}$$

where $2R_2 \sin(\alpha) \pi R_2 \cdot d\alpha$ is the surface area of the red ring, $\rho_2$ is the surface charge density on the right sphere and $Q_2 = 4R_2^2 \pi$ is the total charge of the right sphere.

Finally, the interaction energy between the left and right sphere is:

$$E = \int_0^\pi E(\alpha) d\alpha = A \int_0^\pi \frac{\sin(\alpha)}{\sqrt{R_2^2 + Z^2 - 2ZR_2 \cos(\alpha)}} e^{-\sqrt{R_2^2 + Z^2 - 2ZR_2 \cos(\alpha)}/\lambda_D} d\alpha \tag{5}$$

where $A = \frac{k_e Q_1 \lambda_D}{\varepsilon_r R_1} sinh\left(\frac{R_1}{\lambda_D}\right) \frac{Q_2}{2}$.

Let us do the following substitution in the integral: $u = \cos(\alpha)$.
Thus in Eq.5 $\sin(\alpha) d\alpha$ can be substituted by $-du$ and we get

$$E = A \int_{-1}^{1} \frac{1}{\sqrt{R_2^2 + Z^2 - 2ZR_2 u}} e^{-\sqrt{R_2^2 + Z^2 - 2ZR_2 u}/\lambda_D} du \tag{6}$$

Finally, let us do this substitution in Eq.6 : $w = -\sqrt{R_2^2 + Z^2 - 2ZR_2 u}/\lambda_D$ and thus

$dw = \frac{ZR_2}{\lambda_D \sqrt{R_2^2 + Z^2 - 2ZR_2 u}} du$ and we get

$$E = \frac{A\lambda_D}{Z \cdot R_2} \int_{w(u=-1)}^{w(u=1)} e^w dw = \frac{A\lambda_D}{Z \cdot R_2} [e^w]_{w(u=-1)}^{w(u=1)} \tag{7}$$

where

$$w(u = -1) = -\frac{\sqrt{R_2^2 + Z^2 + 2ZR_2}}{\lambda_D} = -\frac{(Z+R_2)}{\lambda_D}, \tag{8}$$

while in the case of $Z > R_2$

$$w(u=1) = -\frac{\sqrt{R_2^2+Z^2-2ZR_2}}{\lambda_D} = -\frac{\sqrt{(Z-R_2)^2}}{\lambda_D} = -\frac{(Z-R_2)}{\lambda_D}, \tag{9}$$

Thus from Eqs.7-9 we get:

$$E(Z) = \frac{A\lambda_D}{ZR_2}\left[e^{-(Z-R_2)/\lambda_D} - e^{-(Z+R_2)/\lambda_D}\right] = \frac{A\lambda_D}{ZR_2} e^{-\frac{Z}{\lambda_D}} \cdot 2\sinh\left(\frac{R_2}{\lambda_D}\right) =$$

$$\frac{k_e Q_1 Q_2 \lambda_D^2}{\varepsilon_r R_1 R_2 Z} \sinh\left(\frac{R_1}{\lambda_D}\right) \sinh\left(\frac{R_2}{\lambda_D}\right) e^{-\frac{Z}{\lambda_D}} \tag{10}$$

where $Z > R_1 + R_2$.

## Results

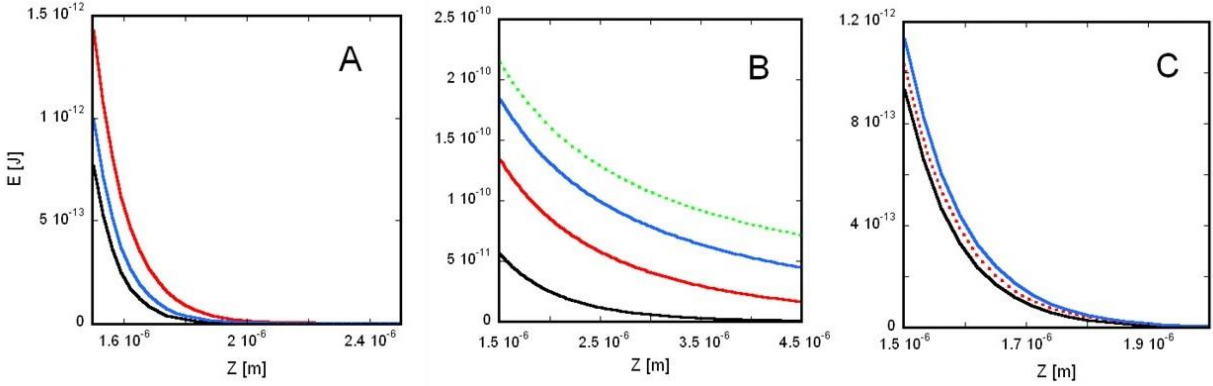

**Figure 2.** *Interaction energy of two charged spheres surrounded by electrolyte (dependence from electrolyte's ion concentration and temperature)*

The smaller sphere with radius $R_1 = 5 \cdot 10^{-7} m$ is located to the left from the larger sphere with radius $R_2 = 10^{-6} m$ (Figure 1). The interaction energy between the two spheres, $E$ is calculated by Eq.10 and plotted against the distance between the centers of the two spheres, $Z (\geq R_1 + R_2 = 1.5 \cdot 10^{-6} m)$. A) The ion concentration, $C$ of the electrolyte (and the respective Debye length from Eq.A2) is: red curve: $0.007 mol/m^3$ ($\lambda_D = 1.15 \cdot 10^{-7} m$); blue curve: $0.01 mol/m^3$ ($\lambda_D = 9.62 \cdot 10^{-8} m$); black curve: $0.013 mol/m^3$ ($\lambda_D = 8.44 \cdot 10^{-8} m$) and the temperature in the case of each curve is $T = 300K$. B) The ion concentration of the electrolyte is: green dotted curve: $0\ mol/m^3$ ($\lambda_D = \infty\ m$); blue curve: $0.000001 mol/m^3$ ($\lambda_D = 9.62 \cdot 10^{-6} m$); red curve: $0.00001 mol/m^3$ ($\lambda_D = 3.04 \cdot 10^{-6} m$); black curve: $0.0001 mol/m^3$ ($\lambda_D = 9.62 \cdot 10^{-7} m$) and the temperature in the case of each curve is $T = 300K$. C) The system's temperature (and the respective Debye length) is: blue curve: $340K$ ($\lambda_D = 1.02 \cdot 10^{-7} m$); red dotted curve: $310K$ ($\lambda_D = 9.78 \cdot 10^{-8} m$); black curve: $280K$ ($\lambda_D = 9.30 \cdot 10^{-8} m$) and the electrolyte's ion concentration in the case of each curve is $0.01 mol/m^3$. In the case of our calculations the surface charge density of each charged sphere is $\rho_s = -0.266\ C/m^2$.

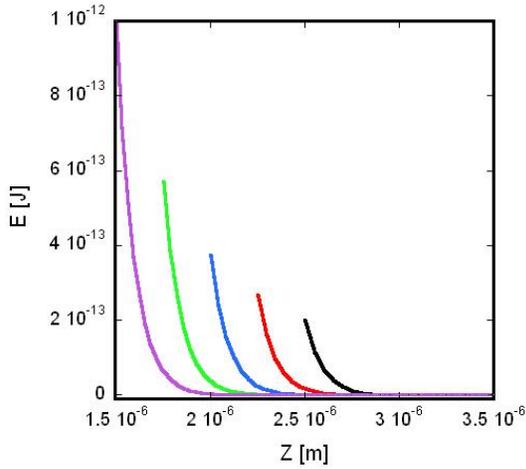

**Figure 3.** *Interaction energy of two charged spheres surrounded by electrolyte (dependence from radius)*

The interaction energy between the two spheres, $E$ is plotted against the distance between the centers of the two spheres, $Z$. The total charge of the left and right sphere is $Q_1 = -8.3566 \cdot 10^{-13} C$ and $Q_2 = -3.34265 \cdot 10^{-12} C$, respectively. The radius of the right sphere (see Figure 1) is $R_2 = 10^{-6} m$, the electrolyte's ion concentration is $C = 0.01 mol/m^3$, the temperature is $T = 300 K$ and the respective Debye length (calculated by Eq.A2) is $\lambda_D = 9.62 \cdot 10^{-8} m$. Purple curve: $R_1 = 5 \cdot 10^{-7} m$; green curve: $R_1 = 7.5 \cdot 10^{-7} m$; blue curve: $R_1 = 1 \cdot 10^{-6} m$; red curve: $R_1 = 1.25 \cdot 10^{-6} m$; black curve: $R_1 = 1.5 \cdot 10^{-6} m$.

## Discussion

Here, by using the recently generalized Shell Theorem [1], Eq.10 is derived to calculate the electric interaction energy between two charged spheres surrounded by electrolyte. Because of the increased screening effect of the electrolyte's ions (i.e. with decreasing Debye length), at any given $Z$ distance between the spheres, the interaction energy decreases with increasing electrolyte ion concentration (Figure 2A,B). The primary reason of this decrease is that the last factor of Eq.10 ( $e^{-\frac{Z}{\lambda_D}}$ ) at a given $Z$ fast decreasing when the Debye length, $\lambda_D$ decreases because of the increasing electrolyte ion concentration (see Figure legends 2A,B). On the other hand $\lambda_D$ increases with increasing temperature (see Eq.A2) and thus at a given $Z$ factor $e^{-\frac{Z}{\lambda_D}}$ increases too causing the increase of the interaction energy between two charged spheres (see Figure legend 2C).

By increasing the radius $R_1$ the $E$ vs. $Z$ curves are shifting to the right (see Figure 3) because the lowest value of $Z(Z_{min} = R_1 + R_2)$ increases. Also at $Z_{min}$ the electric interaction energy $E(Z_{min})$ is getting smaller. This is the case because with increasing $R_1$ the distance between the charges of the spheres are increasing and thus the screening effect of the electrolyte's ions increases too.

By using Eq.10 one can calculate the electric interaction energy between two charged spheres surrounded inside and outside by electrolyte. This equation is a generalization of the Coulomb equation (for charge-charge interaction in vacuum [6]). One can get from Eq.10 the Coulomb equation by taking infinite long Debye length (that is characteristic for vacuum):

$$E(Z) = \frac{k_e Q_1 Q_2}{\varepsilon_r Z}\left\{\lim_{\lambda_D \to \infty} e^{-\frac{Z}{\lambda_D}} \frac{\lambda_D}{R_1} \sinh\left(\frac{R_1}{\lambda_D}\right) \frac{\lambda_D}{R_2} \sinh\left(\frac{R_2}{\lambda_D}\right)\right\} = \frac{k_e Q_1 Q_2}{\varepsilon_r Z}\left\{\lim_{\lambda_D \to \infty} e^{-\frac{Z}{\lambda_D}} \frac{\lambda_D}{R_1}\left[\frac{R_1}{\lambda_D} + \frac{1}{3!}\left(\frac{R_1}{\lambda_D}\right)^3 + \frac{1}{5!}\left(\frac{R_1}{\lambda_D}\right)^5 + \cdots\right]\right\}\left\{\lim_{\lambda_D \to \infty} \frac{\lambda_D}{R_2}\left[\frac{R_2}{\lambda_D} + \frac{1}{3!}\left(\frac{R_2}{\lambda_D}\right)^3 + \frac{1}{5!}\left(\frac{R_2}{\lambda_D}\right)^5 + \cdots\right]\right\} = \frac{k_e Q_1 Q_2}{\varepsilon_r Z} \quad (11)$$

By using Eq.10 one can also calculate the total electric interaction energy of several separated charged spheres surrounded inside and outside by electrolyte:

$$E = \sum_{i=1}^{N} \sum_{\substack{j=1 \\ j \neq i}}^{N} \frac{k_e Q_i Q_j \lambda_D^2}{\varepsilon_r R_i R_j Z_{ij}} \sinh\left(\frac{R_i}{\lambda_D}\right) \sinh\left(\frac{R_j}{\lambda_D}\right) e^{-\frac{Z_{ij}}{\lambda_D}} \quad (12)$$

where $N$ is the number of spheres, $Q_i$, $R_i$, is the total charge and radius of the $i$-th sphere, respectively and $Z_{ij}$ (where $Z_{ij} > R_i + R_j$ ) is the distance between the centers of the $i$-th and $j$-th sphere.

## Conclusions

By using the recently generalized version of Newton's Shell Theorem [1] analytical equations are derived to calculate the electric interaction energy between two separated charged spheres surrounded outside and inside by electrolyte. This electric interaction energy is calculated as a function of the electrolyte's ion concentration, temperature, distance between the spheres and size of the spheres. At the same distance the absolute value of the interaction energy decreases with increasing electrolyte ion concentration and increases with increasing temperature. At zero electrolyte ion concentration the derived analytical equation transforms into the Coulomb equation. Finally, the analytical equation is generalized to calculate the electric interaction energy of N separated charged spheres surrounded by electrolyte.

## Acknowledgement

The author is very thankful for Chinmoy Kumar Ghose.

## *Appendix 1*

The Debye length in an electrolyte is calculated by [7]

$$\lambda_D = \left(\frac{\varepsilon_0 \varepsilon k_B T}{e^2 N_a \sum_{j=1}^{N} c_j^0 q_j^2}\right)^{\frac{1}{2}} \tag{A1}$$

where $\varepsilon_0 = 8.85 \cdot 10^{-12} C^2 J^{-1} m^{-1}$ is the vacuum permittivity, $\varepsilon$ is the relative static permittivity of the electrolyte, $k_B = 1.38 \cdot 10^{-23} J K^{-1}$ is the Boltzmann constant, $T$ is the absolute temperature, $e = 1.6 \cdot 10^{-19} C$ is the charge of a positive monovalent ion, $N_a = 6 \cdot 10^{23} mol^{-1}$ is the Avogadro's number, $c_j^0 \; mol/m^3$ is the mean concentration of the j-th species of ions in the electrolyte, $q_j$ is the number of elementary charges in an ion of the j-th species (e.g. in the case of bivalent ions $q_j = 2$). In this paper we consider overall neutral electrolytes containing only monovalent positive and negative ions of the same concentration, $C$. In this case Eq.A1 is simplified to:

$$\lambda_D = \left(\frac{\varepsilon_0 \varepsilon k_B T}{e^2 N_a 2C}\right)^{\frac{1}{2}} \tag{A2}$$